\documentclass[letterpaper,twocolumn]{article} 
\usepackage{times}  
\usepackage{helvet}  
\usepackage{courier}
\usepackage[hyphens]{url} 
\usepackage{graphicx}
\usepackage{caption} 
\usepackage{algorithm}
\usepackage{algorithmic}
\usepackage{newfloat}
\usepackage{listings}
\usepackage{geometry}
\floatstyle{ruled}
\newfloat{listing}{tb}{lst}{}
\floatname{listing}{Listing}
\usepackage[utf8]{inputenc}
\usepackage[english]{babel}
\usepackage{enumerate}
\usepackage{ amssymb}
\usepackage{amsmath}
\usepackage{amsthm}
\usepackage{ dsfont }
\usepackage{xspace}
\usepackage{subcaption}
\usepackage{etoolbox}
\def\msk{\medskip}
\def\ssk{\smallskip}
\def\pagebreak{\vfill\eject}
\def\leftdisplay#1$${\leftline{\hskip 36pt$\displaystyle{#1}$}$$}

\newcount\commenton   
\def\comment#1 \par{\ifnum\commenton=1
    \medskip{\it [#1]}\par\medskip\fi}
\newcount\commentaon       
\def\commenta#1 \par{\ifnum\commentaon=1
    \medskip{\it [#1]}\par\medskip\fi}
\newcount\commentbon
\def\commentb#1 \par{\ifnum\commentbon=1
    \medskip{\it [#1]}\par\medskip\fi}

\newcount\commentcon
\def\commentc#1 \par{\ifnum\commentcon=1
    \medskip{#1}\par\medskip\fi}

   \newcount\commentpion 
\def\commentpi#1 \par{\ifnum\commentpion=1
    {{#1}}\par\fi}
\def\commentpistart#1 \par{\ifnum\commentpion=1 
    \medskip{\noindent {\em Proof:~}}{{#1}}\par\fi}
\def\commentpiend#1 \par{\ifnum\commentpion=1 
    {{#1}}\hfill $\Box$\par\medskip\fi}
\def\commentpistartend#1 \par{\ifnum\commentpion=1 
    \medskip{\noindent {\em Proof:~}}{{#1}}\hfill $\Box$\par\medskip\fi}

    \newcount\commentphon 
\def\commentph#1 \par{\ifnum\commentphon=1
    {{#1}}\par\fi}
\def\commentphstart#1 \par{\ifnum\commentphon=1 
    \medskip{\noindent {\em Proof:~}}{{#1}}\par\fi}
\def\commentphend#1 \par{\ifnum\commentphon=1 
    {{#1}}\hfill $\Box$\par\medskip\fi}
\def\commentphstartend#1 \par{\ifnum\commentphon=1 
    \medskip{\noindent {\em Proof:~}}{{#1}}\hfill $\Box$\par\medskip\fi}

\def\summ_#1{\setbox1=\hbox{$\displaystyle\sum_{#1}$}
             \setbox2=\hbox{$\displaystyle\sum$}
             \dimen1=0.25\wd1
             \dimen2=0.25\wd2
             \advance \dimen1 by -\dimen2
             \dimen3 = 0.5\dimen1
             \hskip -\dimen3
             \hbox {$\displaystyle\sum_{#1}$}
             \hskip -\dimen1
             }

\def\Nats{{\hbox{$\mathpalette{}{I\kern-.33em N}$}}}
\def\Reals{{\hbox{$\mathpalette{}{I\kern-.33em R}$}}}

\newcommand{\set}[1]{{\{#1\}}}
\newcommand{\sth}{\, : \,}    

\newcommand{\emp}{\emptyset}

\newcommand{\calA}{{\cal A}}

\newcommand{\calC}{{\cal C}}

\newcommand{\calH}{{\cal H}}
\newcommand{\calI}{{\cal I}}

\newcommand{\calL}{{\cal L}}
\newcommand{\calM}{{\cal M}}

\newcommand{\calO}{{\cal O}}
\newcommand{\calP}{{\cal P}}

\newcommand{\calS}{{\cal S}}

\newtheorem{proposition}{Proposition}
\newtheorem{lemma}{Lemma}
\newtheorem{theorem}{Theorem}

\newtheorem{definition}{Definition}

\def\ppp{{\sf P}}
\def\np{{\sf NP}}
\def\conp{{\sf coNP}}

\newcommand{\succceq}{\succcurlyeq} 
\newcommand{\precceq}{\preccurlyeq} 

\newcommand{\la}{\langle}
\newcommand{\ra}{\rangle}

\newcommand{\Qplus}{{\mathbb{Q}^+}}

\newcommand{\Cons}{\sigma}

\newcommand{\vphi}{\varphi}
\newcommand{\LA}{\calL^\calA}
\newcommand{\Lstrict}{\calL_{<}^\calA}
\newcommand{\Lnonstrict}{\calL_{\le}^\calA}

\newcommand{\cardup}{t}

\newcommand{\modelsH}{\models} 

\newcommand{\modelsoplus}{\models^\oplus}

\newcommand{\calCone}{\calC(1)}
\newcommand{\calCt}{\calC(\cardup)}
\newcommand{\calCequiv}{\calC(\equiv)}
\newcommand\modelsCone{\models_{\calCone}}

\newcommand\modelsCoplust{\models_{\calCt}^\oplus}
\newcommand\modelsCoplusequiv{\models_{\calCequiv}^\oplus}

\newcommand\modelsCoplustwo{\models_{\calC(2)}^\oplus}
\newcommand\modelsCoplusthree{\models_{\calC(3)}^\oplus}

\newcommand\modelsCprimeone{\models_{C'(1)}}

\newcommand\calConestar{\calC(1^*)}
\newcommand\modelsConestar{\models_{\calConestar}}

\newcommand{\Suppphi}{\textit{Supp}^\varphi}
\newcommand{\Oppphi}{\textit{Opp}^\varphi}
\newcommand{\Indphi}{\textit{Ind}^\varphi}
\newcommand{\SuppphiC}{\textit{Supp}_\calC^\varphi}
\newcommand{\OppphiC}{\textit{Opp}_\calC^\varphi}
\newcommand{\IndphiC}{\textit{Ind}_\calC^\varphi}

\newcommand{\MIB}{\textit{MIB}}

\newcommand{\Supp}{\textit{Supp}}
\newcommand{\Opp}{\textit{Opp}}
\newcommand{\Gammadestrict}{\Gamma^{(\le)}}

\newcommand{\one}{1}

\title{Computation and Complexity of Preference Inference Based on Hierarchical Models
\\
\textit{Longer Version of IJCAI'15 publication}
}
\author{Nic Wilson, Anne-Marie George and Barry O'Sullivan \\
Insight Centre for Data Analytics, 
School of Computer Science and IT \\
University College Cork, Ireland \\
\{nic.wilson, annemarie.george, barry.osullivan\}@insight-centre.org
}
\date{}

\begin{document}

\maketitle

\begin{abstract}
 Preference Inference involves inferring additional user preferences from elicited or observed preferences,
based on
assumptions regarding
the form of the user's preference relation.
In this paper we consider a situation in which
alternatives have an associated vector of costs, each component corresponding to a different criterion,
and are compared using a kind of lexicographic order, similar to the way alternatives are compared in a Hierarchical Constraint Logic Programming model.
It is assumed that the user has some (unknown) importance ordering on criteria, and that
to compare two alternatives, firstly, the combined cost of each alternative with respect to the most important criteria are compared;
only if these combined costs are equal, are the next most important criteria considered.
The preference inference problem then consists of determining whether a preference statement can be inferred from a set of input preferences. We show that this problem is
\conp-complete,
even if one restricts the cardinality of the equal-importance sets to have at most two elements, and one only considers non-strict preferences.
However, it is polynomial if it is assumed that the user's ordering of criteria is a total ordering;
it is also polynomial if the sets of equally important criteria are all equivalence classes of a given fixed equivalence relation.
We give an efficient polynomial algorithm for these cases, which also throws light on the structure of the inference.
\end{abstract}

\section{Introduction}

There are increasing opportunities for decision making/support systems to take into account the preferences of individual users, with the user preferences being elicited or
observed
from the user's behaviour.
However, users tend to have limited patience for preference elicitation,
so such a system will tend to have a very incomplete picture of the user preferences.
\emph{Preference Inference} involves inferring additional user preferences from elicited or observed preferences,
based on assumptions regarding the form of the user's preference relation.
More specifically, given a set of input preferences $\Gamma$,
and a set of preference models $\calM$ (considered as candidates for the user's preference model),
we infer a preference statement $\vphi$ if
every model in $\calM$ that satisfies $\Gamma$ also satisfies $\vphi$.
Preference Inference can take many forms, depending on the choice of $\calM$,
and on the choices of language(s) for the input and inferred statements.
For instance, if we just assume that the user model is a total order (or total pre-order),
we can set $\calM$ as the set of total [pre-]orders over a set of alternatives.
This leads to a relatively cautious form of inference (based on transitive closure),
including, for instance, the dominance relation for CP-nets and some related systems, e.g.,
\cite{cpjair04,tcpJAIR06,Bouveret-etal-IJCAI09,BienvenuLangWilson10}.

Often it can be valuable to obtain a much less cautious form of inference, for example, in order to have some help in deciding which options to show to the user next in a recommender system \cite{BridgRicci07,TrabelsiWilsonBridgeRicciIJAIT11}.
This includes assuming that the user's preference relation in a multi-objective context is based on a simple weighted sum of objectives
(as in a simple form of a Multi-Attribute Utility Theory model \cite{MCDA2005});
this kind of preference inference is considered in
\cite{BridgRicci07,MarinescuRazakWilsonCP2013};
or alternatively, assuming different lexicographic forms for the preference models
as in \cite{WilsonIJCAI09,TrabelsiWilsonBridgeRicciIJAIT11,WilsonECAI14}.
Note that all these systems involve reasoning about what holds in a set of preference models. This contrasts with work in preference learning
\cite{Fuern-Hueller-Preference-Learning,Dombi-Learning-Lex-EJOR-2007,FlachM07-Lex-ranker,Huellermeier-PL-12-Learn-Lex-Trees,BoothCLMS-ECAI10}
 that typically learns a single model.

In this paper we consider a situation in which
%
%
alternatives have an associated vector of costs, each component corresponding to a different criterion,
and are compared using a kind of lexicographic order, similar to the way alternatives are compared in a Hierarchical Constraint Logic Programming (HCLP) model
\cite{WilsonBorning1993}.
It is assumed that the user has some (unknown) importance ordering on criteria, and that
to compare two alternatives, firstly, the combined cost of each alternative with respect to the most important criteria are compared;
only if these combined costs are equal, are the next most important criteria considered.

We consider the case where the
input preference statements are of a simple form that one alternative is preferred to another alternative,
 where
 we allow the expression of both strict and non-strict preferences
 (in contrast to most related preference logics, such as  \cite{WilsonAIJ11,cpjair04,WilsonIJCAI09,WilsonECAI14}).
This form of preference is natural in many contexts, including for conversational recommender systems \cite{BridgRicci07}.
The preference inference problem then consists of determining whether a preference statement can be inferred from a set of input preferences,
i.e., if every model (of the assumed form) satisfying the inputs also satisfies the query.
We show that this problem is \conp-complete, even if one restricts the cardinality of the equal-importance sets to have at most two elements, and one only considers non-strict preferences.
However, it is polynomial if it is assumed that the user's ordering of criteria is a total ordering;
it is also polynomial if the sets of equally important criteria are all equivalence classes of a given fixed equivalence relation.
We give an efficient polynomial algorithm for these cases, which also throws light on the structure of the inference.

\commenta
**
The most closely connected work is that in Wilson-2014.

\commenta
Don't forget to say that it's notable that language can express both strict and non-strict preferences.

Section~\ref{sec:PrefLogic} defines our simple preference logic based on hierarchical models,
along with some associated preference inference problems.
Section~\ref{sec:conp} shows that in general the preference inference problem is \conp-complete.
Section~\ref{sec:singleton-sequence} considers the case where the importance ordering on criteria is a total order, and gives a polynomial algorithm for consistency.
Section~\ref{sec:Conclusions} concludes.

\commenta
** Note the below:

\ssk

\section{A Preference Logic Based on Hierarchical Models}
\label{sec:PrefLogic}

We consider preference models, based on an importance ordering of criteria, that is basically lexicographic, but involving a combination of criteria which are at the same level in the importance ordering.
We call these ``HCLP models", because
models of a similar kind appear in the HCLP system \cite{WilsonBorning1993}
  (though we have abstracted away some details from the latter system).

\commenta
Monotonicity of $\oplus$ probably not necessary for results,
except that our hardness proof assumes $\one \oplus \one\not=0$.

\paragraph*{HCLP structures:}
Define an HCLP structure to be a tuple
$\calS = \la \calA, \oplus, \calC \ra$,
where $\calA$ (the set of alternatives) is a finite set;
$\oplus$ is an
associative, commutative and monotonic operation
($x \oplus y \le z \oplus y$ if $x \le z$) on
the non-negative rational numbers $\Qplus$, with
identity element $0$;
and $\calC$ (known as the set of ($\calA$-)evaluations)
is a set of functions from $\calA$ to $\Qplus$.
We also assume that
operation
$\oplus$ can be computed in linear time (which holds for natural definitions of $\oplus$, including addition and max).
The evaluations in $\calC$ may be considered as representing criteria or objectives under which the alternatives are evaluated.
For $c\in\calC$ and $\alpha\in\calA$,
if $c(\alpha) = 0$ then $\alpha$ fully satisfies the objective corresponding to $c$; more generally, the smaller the value of $c(\alpha)$, the better $\alpha$ satisfies the $c$-objective.

With each subset $C$ of $\calC$ we define ordering $\precceq_C^\oplus$ on $\calA$ by
$\alpha \precceq_C^\oplus \beta$ if and only if
$\bigoplus_{c \in C} c(\alpha) \le \bigoplus_{c \in C} c(\beta)$.
Relation $\precceq_C^\oplus$ represents how well the alternatives satisfy the multi-set of evaluations $C$ if the latter are considered equally important.
$\precceq_C^\oplus$ is a total pre-order (a weak order, i.e., a transitive and complete binary relation).
If $\alpha \precceq_C^\oplus \beta$, we might also write
$\beta \succceq_C^\oplus \alpha$.
We write $\equiv_C^\oplus$ for the associated equivalence relation on $\calA$,
given by $\alpha \equiv_C^\oplus \beta$ $\iff$ $\alpha \precceq_C^\oplus \beta$
and $\beta \precceq_C^\oplus \alpha$.
We write $\prec_C^\oplus$ for the associated strict weak ordering,
defined by $\alpha \prec_C^\oplus \beta$ $\iff$
$\alpha \precceq_C^\oplus \beta$ and $\beta \not\precceq_C^\oplus \alpha$.
Thus,
$\alpha \equiv_C^\oplus \beta$ if and only if
$\bigoplus_{c \in C} c(\alpha) = \bigoplus_{c \in C} c(\beta)$;
and
$\alpha \prec_C^\oplus \beta$ if and only if
$\bigoplus_{c \in C} c(\alpha) < \bigoplus_{c \in C} c(\beta)$.

\paragraph{HCLP models:}
An HCLP model $H$ based on $\la \calA, \oplus, \calC \ra$
is defined to be an ordered partition
$(C_1, \ldots, C_k)$ of a (possibly empty) subset $\Cons(H)$
of $\calC$.
The sets $C_i$ are called the \emph{levels of $H$},
which are thus non-empty, disjoint  and have union $\Cons(H)$.
If $c \in C_i$ and $c' \in C_j$, and $i < j$,
then we say that \emph{$c$ appears before $c'$}
(and \emph{$c'$ appears after $c$}) in $H$.
Associated with $H$ is an ordering relation $\precceq^\oplus_H$ on $\calA$ given by:

$\alpha \precceq^\oplus_H \beta$ if and only if either:
\begin{itemize}
  \item[(I)]  for all $i=1, \ldots, k$,
$\alpha \equiv^\oplus_{C_i} \beta$; or
  \item[(II)] there exists some $i \in\set{1, \ldots, k}$ such that
  (i) $\alpha \prec^\oplus_{C_i} \beta$ and (ii) for all $j$ with $1 \le j < i$,  $\alpha \equiv^\oplus_{C_j} \beta$.
\end{itemize}
Relation $\precceq^\oplus_H$ is a kind of lexicographic order on $\calA$,
where the multi-set $C_i$ of evaluations at the same level are first combined into a single evaluation.
$\precceq^\oplus_H$
is a weak order on $\calA$.
We write $\equiv^\oplus_H$ for the associated equivalence relation
(corresponding with condition (I)),
and $\prec^\oplus_H$ for the associated strict weak order
(corresponding with condition (II)),
so that $\precceq^\oplus_H$ is the disjoint union of
$\prec^\oplus_H$ and $\equiv^\oplus_H$.
If $\Cons(H) = \emp$ then
then the first condition for $\alpha \precceq^\oplus_H \beta$ holds vacuously
(since $k=0$),
so we have $\alpha \precceq^\oplus_H \beta$
for all $\alpha, \beta\in\calA$,
and $\prec^\oplus_H$ is the empty relation.

\paragraph*{Preference Language Inputs:}
Let $\calA$ be a set of alternatives.
We define $\Lnonstrict$ to be the set of statements of the form $\alpha \le \beta$, for $\alpha, \beta \in \calA$
(the \emph{non-strict} statements),
we write $\Lstrict$ for the set of statements
of the form $\alpha < \beta$, for $\alpha, \beta \in \calA$
(the \emph{strict} statements),
and let $\LA = \Lnonstrict \cup \Lstrict$.
If $\vphi$ is the preference statement $\alpha \le \beta$
then $\neg\vphi$ is defined to be the preference
statement $\beta < \alpha$.
If $\vphi$ is the preference statement $\alpha < \beta$
then $\neg\vphi$ is defined to be the preference
statement $\beta \le \alpha$.

\paragraph{Satisfaction of preference statements:}
For HCLP model $H$ over HCLP structure $\la \calA, \oplus, \calC \ra$,
we say that $H$ satisfies $\alpha \le \beta$
(written $H\modelsoplus \alpha \le \beta$) if
$\alpha \precceq^\oplus_H \beta$ holds.
Similarly, we say that $H$ satisfies $\alpha < \beta$
(written $H\modelsoplus \alpha < \beta$)
if $\alpha \prec^\oplus_H \beta$.
For $\Gamma\subseteq\LA$,
we say that $H$ satisfies $\Gamma$
(written $H\modelsoplus\Gamma$)
if $H$ satisfies $\vphi$ for all $\vphi\in\Gamma$.

\commentph
\begin{lemma}
\label{le:basic-negation}
Let $H$ be a  HCLP model over HCLP structure $\calS$.
Then,
$H$ satisfies $\vphi$
if and only if
$H$ does not satisfy $\neg\vphi$.
\end{lemma}

\commentphstartend
Write $\calS$ as $\la \calA, \oplus, \calC \ra$.
It is sufficient to show that, for any $\alpha, \beta\in\calA$
$H$ satisfies $\alpha \le \beta$
if and only if $H$ does not satisfy $\beta < \alpha$.
We have that $H$ satisfies $\alpha \le \beta$
if and only if
$\alpha \precceq^\oplus_H \beta$,
which, since $\precceq^\oplus_H$ is a weak order, is
if and only if
$\beta \not\prec^\oplus_H \alpha$,
i.e., $H$ does not satisfy $\beta < \alpha$.


\paragraph{Preference Inference/Deduction relation:}
We are interested in different restrictions on the set of models,
and the corresponding inference relations.
Let $\calM$ be a set of HCLP models over HCLP structure $\la \calA, \oplus, \calC \ra$.
For $\Gamma \subseteq \LA$, and $\vphi \in \LA$,
we say that
$\Gamma \models_\calM^\oplus \vphi$,
if $H$ satisfies $\vphi$ for every $H \in\calM$ satisfying
$\Gamma$.
Thus, if we elicit
some
preference statements $\Gamma$
of a user, and we assume that their preference relation is an HCLP model in $\calM$ (based on the HCLP structure),
then  $\Gamma \models_\calM^\oplus \vphi$ holds if and only if we can deduce
that the user's HCLP model $H$ satisfies $\vphi$.

\paragraph{Consistency:}
For set of HCLP models $\calM$ over HCLP structure $\la \calA, \oplus, \calC \ra$, and set of preference statements $\Gamma\subseteq\LA$,
we say that
$\Gamma$ is $(\calM,\oplus)$-consistent if there exists $H \in \calM$
such that $H \modelsoplus \Gamma$;
otherwise, we say that $\Gamma$ is $(\calM,\oplus)$-inconsistent.
In the usual way, because of the existence of a negation operator,
deduction can be reduced to checking (in)consistency.

\begin{proposition}
\label{pr:basic-cons-deduction}
$\Gamma \models_\calM^\oplus \vphi$ if and only if
$\Gamma\cup\set{\neg\vphi}$ is $(\calM,\oplus)$-inconsistent.
\end{proposition}

\commentpistart
Suppose that $\Gamma \models_\calM^\oplus \vphi$.
By definition, $H$ satisfies $\vphi$ for every $H \in\calM$ satisfying (every element of) $\Gamma$.
Thus, there exists no $H \in\calM$ that satisfies $\Gamma$ and $\neg \vphi$,
which implies that $\Gamma\cup\set{\neg\vphi}$ is $(\calM,\oplus)$-inconsistent.

\commentpiend
Conversely, suppose $\Gamma\cup\set{\neg\vphi}$ is $(\calM,\oplus)$-inconsistent. By definition, there exists no $H \in\calM$ that satisfies $\Gamma \cup \neg \vphi$.
Thus, every $H \in\calM$ that satisfies $\Gamma$ does not satisfy $\neg\vphi$,
and therefore satisfies $\vphi$. Hence, $\Gamma \models_\calM^\oplus \vphi$.

Let $\cardup$ be some
number in $\set{1, 2, \ldots, |\calC|}$.
We define $\calCt$ to be the set of
all  HCLP models $(C_1, \ldots, C_k)$
based on
 HCLP structure $\la \calA, \oplus, \calC \ra$
such that $|C_i| \le \cardup$, for all $i = 1,\ldots, k$.
An element of $\calCone$ thus corresponds to
a sequence of singleton sets of evaluations;
we identify it with
a sequence of evaluations $(c_1, \ldots, c_k)$ in $\calC$.
Thus,
$\Gamma \modelsCoplust \vphi$ if and only if $H\modelsoplus \vphi$
for all $H\in\calCt$ such that $H\modelsoplus \Gamma$.
Note that for $\cardup=1$, these definitions do not depend on $\oplus$
(since there is no combination of evaluations involved),
so we may drop any mention of $\oplus$.

Let $\equiv$ be an equivalence relation on $\calC$.
We define $\calCequiv$
 to be the set of
all  HCLP models $(C_1, \ldots, C_k)$  such that
each $C_i$ is an equivalence class with respect to $\equiv$.
It is easy to see that the relation
$\modelsCoplusequiv$ is the same as the relation
$\modelsCprimeone$
where $C'$ is defined as follows.
$C'$ is in 1-1 correspondence with the set of $\equiv$-equivalence classes of $\calC$.
If $E$ is the $\equiv$-equivalence class of $\calC$
corresponding with $c' \in C'$
then, for $\alpha\in\calA$,
$c'(\alpha)$ is defined to be $\bigoplus_{c\in E} c(\alpha)$.

For $\models$ either being  $\modelsCoplust$ for some $\cardup \in\set{1, 2, \ldots, |\calC|}$,
or being $\modelsCoplusequiv$ for some equivalence relation $\equiv$ on $\calC$, we consider the following decision problem.

\msk\noindent
{\sc HCLP-Deduction for $\models$}:
Given $\calC$, $\Gamma$ and $\vphi$
is it the case that $\Gamma \models \vphi$?

\msk
In Section~\ref{sec:singleton-sequence},
we will show that this problem is
polynomial for $\models$ being $\modelsCoplust$ when $\cardup = 1$.
Thus it is polynomial also for $\modelsCoplusequiv$,
for any equivalence relation $\equiv$.
It is \conp-complete for
$\models$ being $\modelsCoplust$ when $\cardup > 1$,
as shown below in Section~\ref{sec:conp}.

\begin{theorem}
\label{th:main-complexity-theorem}
{\sc HCLP-Deduction for $\modelsCoplust$}
is polynomial when $\cardup=1$,
and is \conp-complete
for any $\cardup > 1$, even if
we restrict the language to non-strict preference statements.
{\sc HCLP-Deduction for $\modelsCoplusequiv$}
is polynomial for any equivalence relation $\equiv$.
\end{theorem}

\subsubsection*{Example}

We consider an example with
alternatives $\calA = \set{\alpha, \beta, \gamma}$,
using $\oplus$ as $+$ (ordinary addition),
and with evaluations $\calC = \set{c_1, c_2, c_3}$, defined as follows.

\ssk
 $c_1(\alpha) = 0$; \  $c_1(\beta) = 2$; \  $c_1(\gamma) = 1$;

 $c_2(\alpha) = 2$; \  $c_2(\beta) = 0$; \  $c_2(\gamma) = 2$;

 $c_3(\alpha) = 1$; \  $c_3(\beta) = 0$; \  $c_3(\gamma) = 0$.
\ssk

\noindent
Suppose that the user tells us that
they consider $\alpha$ to be at least as good as $\beta$.
We represent this as the non-strict preference $\alpha \le \beta$.
This eliminates some models;
for instance, $\alpha \le \beta$ is not satisfied by model $H$ equalling $(\set{c_1, c_2}, \set{c_3})$,
since $c_1(\alpha) \oplus c_2(\alpha) = 2 =  c_1(\beta) \oplus c_2(\beta)$,
and $c_3(\beta) < c_3(\alpha)$, which implies that
$\beta \prec^\oplus_H \alpha$, and thus $H \models \beta<\alpha$
and $H \not\models \alpha \le \beta$.

Now, $\alpha \le \beta$ does not entail any preference
between $\beta$ and $\gamma$, as
can be seen by considering the two models
$(\set{c_1})$ and $(\set{c_1, c_2})$, each with just a single level,
and both satisfying $\alpha \le \beta$.
Model $(\set{c_1, c_2})$ satisfies
$\beta < \gamma$,
since $c_1(\beta) \oplus c_2(\beta) = 2
< 3 = c_1(\gamma) \oplus c_2(\gamma)$,
whereas model $(\set{c_1})$ satisfies $\gamma < \beta$,
since $c_1(\gamma) < c_1(\beta)$.
However, it can be shown that
 $\alpha \le \beta \modelsCoplusthree \alpha \le \gamma$,
so we infer that $\alpha$ is non-strictly preferred to $\gamma$.

A strict preference for $\alpha$ over $\beta$ entails additional inferences,
for instance,
$\alpha < \beta \modelsCoplusthree \gamma < \beta$.
If we restrict the set of models to $\calCone$
(thus assuming that the importance ordering on $\calC$ is a total order)
we get slightly stronger inferences still, obtaining in addition:
$\alpha < \beta \modelsCone \alpha < \gamma$.

\section{Proving \conp-completeness of HCLP-Deduction for $\modelsCoplust$ for $\cardup > 1$}
\label{sec:conp}

Given an arbitrary 3-SAT instance we will show that we can construct a set $\Gamma$ and a statement $\alpha\le\beta$
such that
the 3-SAT instance has a satisfying truth assignment if and only if
$\Gamma \not\modelsCoplust \alpha\le\beta$
(see Proposition~\ref{pr:NPhard}).
This then implies that determining if $\Gamma \not\modelsCoplust \alpha \le \beta$ holds is \np-hard.

We have that $\Gamma \not\modelsCoplust \alpha\le\beta$
if and only if there exists an HCLP-model $H \in \calCt$ such that
$H \modelsoplus \Gamma$ and $H \not\modelsoplus \alpha\le\beta$.
For any given $H$, checking that
$H \modelsoplus \Gamma$ and $H \not\modelsoplus \alpha\le\beta$
can be performed in polynomial time.
This implies that determining if
$\Gamma \modelsCoplust \alpha\le\beta$ holds is \conp-complete.

\commenta
Perhaps more detail above.

\commentc
$\one$ can be replaced by an non-zero element $x$
(and we could use other scales apart from $\Qplus$).


\newcommand{\numpvars}{r}
\newcommand{\numclauses}{s}

\newcommand{\qiplus}{q_i^{+}}
\newcommand{\qiminus}{q_i^{-}}
\newcommand{\qplus}{q^{+}}
\newcommand{\qminus}{q^{-}}

\msk
Consider an arbitrary 3-SAT instance
based on propositional variables $p_1, \ldots, p_\numpvars$, consisting
of clauses $\Lambda_j$,
for $j= 1, \ldots, \numclauses$.
For each propositional variable $p_i$ we associate
two evaluations $\qiplus$ and $\qiminus$,
where $\qiminus$ corresponds with literal $\neg p_i$
and $\qiplus$ corresponds with literal $p_i$.

The idea behind the construction is as follows:
we generate a (polynomial size) set $\Gamma \subseteq \Lnonstrict$
as the disjoint union of sets $\Gamma_1$, $\Gamma_2$ and $\Gamma_3$.
$\Gamma_1$ is chosen so that if
$H \modelsoplus \Gamma_1$ then,
for each $i =1, \ldots, r$,
 $\Cons(H)$
cannot contain both $\qiplus$ and $\qiminus$,
i.e., $\qiplus$ and $\qiminus$ do not both appear in $H$.
(Recall $H$ is an ordered partition of $\Cons(H)$,
so that $\Cons(H)$ is the subset of $\calC$ that appears in $H$.)
If $H \modelsoplus \Gamma_2$ and $H \modelsoplus \beta<\alpha$
then $\Cons(H)$ contains either $\qiplus$ or $\qiminus$.
Together, this implies that if
$H \modelsoplus \Gamma$ and $H \not\modelsoplus \alpha\le\beta$
then
for each propositional variable $p_i$,
model $H$ involves either   $\qiplus$ or $\qiminus$,
but not both.
$\Gamma_3$ is used to make the correspondence with the clauses.
For instance, if one of the clauses is $p_2 \vee \neg p_5 \vee p_6$
then
any HCLP model $H\in\calCt$ of $\Gamma \cup\set{\beta<\alpha}$ will involve either $\qplus_2$, $\qminus_5$, or $\qplus_6$.

Suppose that $H$ satisfies $\Gamma$ but not $\alpha\le\beta$.
We can generate a satisfying assignment of the 3-SAT instance,
by assigning $p_i$ to $1$ (TRUE) if and only if
$\qiplus$ appears in $H$.

\msk
We describe the construction more formally below.

\paragraph*{Defining $\calA$ and $\calC$:}
The set of alternatives $\calA$ is defined to be:
$\set{\alpha, \beta} \cup \set{\alpha_i, \beta_i, \delta_i \sth i = 1, \ldots, \numpvars} \cup \set{\gamma_i^k \sth i = 1, \ldots, \numpvars, k = 1, \ldots, t-1} \cup\set{\theta_j, \tau_j \sth j = 1, \ldots,\numclauses}$.

We define the set of evaluations $\calC$ to be
$\set{c^*} \cup \set{\qiplus, \qiminus \sth i = 1, \ldots, \numpvars} \cup A_1 \cup \dots \cup A_\numpvars$, where $A_i = \set{a_i^k \sth k = 1, \dots, t-1}$.
Both $\calA$ and $\calC$ are of polynomial size.

\paragraph*{Satisfying $\beta < \alpha$:}
The evaluations on $\alpha$ and $\beta$ are defined as follows:
\begin{itemize}
  \item $c^*(\alpha) = \one$, and for all $c \in \calC-\set{c^*}$, $c(\alpha) = 0$.
  \item For all $c \in \calC$, $c(\beta) = 0$.
\end{itemize}
It immediately follows that:
$H \modelsoplus \beta<\alpha$
$\iff$ $\Cons(H) \ni c^*$.

\paragraph*{The construction of $\Gamma_1$:}
For each $i=1,\ldots, \numpvars$, we
define $\Gamma_1^i = \set{\delta_i\le\gamma_i^k, \ \gamma_i^k \le\delta_i \sth k = 1, \dots, t-1}$,
and we let $\Gamma_1 = \bigcup_{i=1}^\numpvars \Gamma_1^i$.
We make use of auxiliary evaluations $A_i = \set{a_i^1, \dots , a_i^{t-1}}$.
The values of the evaluations on $\gamma_i^k$ and $\delta_i$ are defined as follows:
\begin{itemize}
  \item $a_i^k(\gamma_i^k) = \one$, and for all $c \in\calC-\set{a_i^k}$ we set $c(\gamma_i^k) = 0$.
  \item $\qiplus(\delta_i) = \qiminus(\delta_i) = \one$,
and for other $c \in \calC$,
$c(\delta_i) = 0$.
\end{itemize}
Thus, for any $B\subseteq A_i$, we have $(\bigoplus_{a \in B} a \oplus \qiplus)(\delta_i) =
\bigoplus_{a \in B} a(\delta_i) \oplus \qiplus(\delta_i) = 0 \oplus \dots \oplus 0 \oplus \one = \one$.
Similarly,
$(\bigoplus_{a \in B} a \oplus \qiminus)(\delta_i) = \one$.
Furthermore,
$(\bigoplus_{a \in B} a \oplus \qiplus)(\gamma_i^k)= \one \Leftrightarrow a_i^k \in B$ and $(\bigoplus_{a \in B} a \oplus \qiminus)(\gamma_i^k) = \one \Leftrightarrow a_i^k \in B$.

\begin{lemma}
\label{le:GammaOne}
$H \modelsoplus \Gamma_1^i$
if and only if either
(i)
$\Cons(H)$ does not contain any element in $A_i$ or $\qiplus$ or $\qiminus$,
i.e., $\Cons(H) \cap (A_i \cup \set{\qiplus, \qiminus}) = \emp$;
or (ii) $\set{a_i^1, \dots a_i^{t-1}, \qiplus}$ is a level of $H$,
and $\Cons(H) \not\ni \qiminus$;
or (iii) $\set{a_i^1, \dots a_i^{t-1}, \qiminus}$ is a level of $H$,
and $\Cons(H) \not\ni \qiplus$.
In particular, if $H \modelsoplus \Gamma_1^i$
then $\Cons(H)$ does not contain
both $\qiplus$ and $\qiminus$.
\end{lemma}

\commenta
** I might define
$c$ appears in $H$ before $c'$
to mean either
$c' \notin\Cons(H)$ or $c$ appears before $c'$ in $H$.
(However, this might be confusing.)

This holds because if $a_i^k$ appears in $H$ before $\qiplus$ and $\qiminus$ for any $k = 1, \dots, t-1$,
then we will have that $H\not\modelsoplus \gamma_i^k \le\delta_i$.
If $\qiplus$ or $\qiminus$ appears in $H$ before $a_i^k$ then we will have
$H\not\modelsoplus \delta_i\le\gamma_i^k$.
Also, if $a_i^k$ and $\qiplus$ are in the same level of $H$
and $\qiminus$ appears later in $H$ then we will have
$H \not\modelsoplus \delta_i\le\gamma_i^k$.
Similarly, if $a_i^k$ and $\qiminus$ are in the same level of $H$
and $\qiplus$ appears later in $H$ then we will have
$H \not\modelsoplus \delta_i\le\gamma_i^k$.
Thus, if $\qiplus$ is in $H$, then all $a_i^k \in A_i$ must be in the same level as $\qiplus$.
Also, $\qiminus$ can not appear before or after $\qiplus$ and can't be in the same level either, since a level can contain at most $t$ elements.
Hence, $\qiminus \notin \Cons(H)$. Similarly, if $\qiminus$ is in $H$, then all $a_i^k \in A_i$ must be in the same level and $\qiplus \notin \Cons(H)$.


\paragraph*{The construction of $\Gamma_2$:}

For each $i=1,\ldots, \numpvars$,
define $\vphi_i$ to be $\alpha_i\le\beta_i$.
We let $\Gamma_2 = \set{\vphi_i \sth i = 1, \ldots, \numpvars}$.
The values of the evaluations on $\alpha_i$ and $\beta_i$ are defined as follows.
We define $c^*(\alpha_i) = \one$,
and for all $c \in \calC - \set{c^*}$,
$c(\alpha_i) = 0$.
Define $\qiplus(\beta_i) = \qiminus(\beta_i) = \one$,
and for all $c \in \calC -\set{\qiplus, \qiminus}$,
$c(\beta_i) = 0$.
Thus, similarly to the previous observations for $\Gamma_1$, $(c^* \oplus \qiplus)(\beta_i) = (c^* \oplus \qiminus)(\beta_i) = \one$ and
$(c^* \oplus \qiplus)(\alpha_i) = (c^* \oplus \qiminus)(\alpha_i) = \one$.
Also, $(\qiplus \oplus \qiminus)(\alpha_i) = 0$ and $(\qiplus \oplus \qiminus)(\beta_i) \geq \one$, because of the monotonicity of $\oplus$, and $(c^* \oplus \qiplus \oplus \qiminus)(\alpha_i) = \one$ and $(c^* \oplus \qiplus \oplus \qiminus)(\beta_i) \geq \one$.

The following result easily follows.

\begin{lemma}
\label{le:GammaTwo}
If $\qiplus$ or $\qiminus$ appears before (or in the same level as) $c^*$ in $H$
then $H \modelsoplus \vphi_i$.
If $\Cons(H) \ni c^*$ and
$H \modelsoplus \vphi_i$ then $\Cons(H) \ni \qiplus$ or $\Cons(H) \ni \qiminus$.
\end{lemma}

\paragraph*{The construction of $\Gamma_3$:}
For each $i=1,\ldots, \numpvars$,
define $Q(p_i) = \qiplus$ and $Q(\neg p_i) = \qiminus$.
This defines the function $Q$ over all literals.
Let us write the
$j$th clause as $l_1 \vee  l_2 \vee l_3$ for literals $l_1$, $l_2$ and $l_3$.
Define $Q_j = \set{Q(l_1), Q(l_2), Q(l_3)}$.
For example, if
the $j$th clause were $p_2 \vee \neg p_5 \vee p_6$
then $Q_j = \set{\qplus_2, \qminus_5, \qplus_6}$.
We define $\psi_j$ to be  $\theta_j\le\tau_j$, and
 $\Gamma_3 = \set{\psi_j \sth j = 1, \ldots\numclauses}$.
Define $c^*(\theta_j) = \one$ and
$c(\theta_j) = 0$ for all $c \in \calC - \set{c^*}$.
Define $q(\tau_j) = \one$ for $q \in Q_j$,
and for all other $c$ (i.e., $c \in \calC-Q_j$), define $c(\tau_i) = 0$.
Thus, similarly as before, for $q,q',q'' \in Q_j$, we have
$(c^* \oplus q \oplus q' \oplus q'')(\theta_j) = (c^* \oplus q \oplus q' )(\theta_j)
= (c^* \oplus q)(\theta_j) = c^*(\theta_j) = \one$ and
$(c^* \oplus q \oplus q' \oplus q'')(\tau_j) \geq (c^* \oplus q \oplus q' )(\tau_j)
\geq (c^* \oplus q)(\tau_j) = \one$.
Also, $(q \oplus q' \oplus q'')(\theta_j) = (q \oplus q' )(\theta_j) = q(\theta_j) = 0$ and
$(q \oplus q' \oplus q'')(\tau_j) \geq (q \oplus q' )(\tau_j) \geq q(\tau_j) = \one$,
because of the monotonicity of $\oplus$, for $q,q',q'' \in Q_j$.
This leads to the following properties.

\begin{lemma}
\label{le:GammaThree}
If some element of $Q_j$ appears in $H$ before $c^*$
then $H \modelsoplus \psi_j$.
If $\Cons(H) \ni c^*$ and
$H \modelsoplus \psi_j$ then
$\Cons(H)$ contains some element of $Q_j$.
\end{lemma}

We set $\Gamma = \Gamma_1 \cup \Gamma_2 \cup \Gamma_3$.
The following result implies that the HCLP deduction problem is \conp-hard (even if we restrict to the case when $\Gamma \cup\set{\vphi} \subseteq \Lnonstrict$).

\begin{proposition}
\label{pr:NPhard}
Using the notation defined above,
the 3-SAT instance is satisfiable
if and only if
$\Gamma \not\modelsCoplust \alpha\le\beta$.
\end{proposition}

\commentpistart
First let us assume that $\Gamma \not\modelsCoplust \alpha\le\beta$.
Then by definition, there exists
an HCLP model $H\in\calCt$ with
$H \modelsoplus \Gamma$ and $H \not\modelsoplus \alpha\le\beta$.
Since $H \not\modelsoplus \alpha\le\beta$ $\iff$ $H \modelsoplus \beta<\alpha$,
we have  $H \modelsoplus \Gamma \cup\set{\beta<\alpha}$.
Because $H \modelsoplus \beta<\alpha$,
$\Cons(H) \ni c^*$.

\commentpi
Because also $H \modelsoplus \Gamma_2^i$,
either $\Cons(H) \ni \qiplus$ or $\Cons(H) \ni \qiminus$,
by Lemma~\ref{le:GammaTwo}.
Since $H \modelsoplus \Gamma_1^i$,
the set $\Cons(H)$ does not contain both $\qiplus$ and $\qiminus$,
by Lemma~\ref{le:GammaOne}.

\commentpi
Let us define a truth function $f: \calP \to \set{0, 1}$
as follows:
$f(p_i) = 1$ $\iff$ $\Cons(H) \ni \qiplus$.
Since $\Cons(H)$ contains exactly one of
$\qiplus$ and $\qiminus$,
we have
$f(p_i) = 0$ $\iff$ $\Cons(H) \ni \qiminus$.
We extend $f$ to negative literals in the obvious way:
$f(\neg p_i) = 1 - f(p_i)$.

\commentpi
Since $H \modelsoplus \Gamma_3 \cup\set{\beta<\alpha}$,
$\Cons(H)$ contains at least one element of each $Q_j$,
by Lemma~\ref{le:GammaThree}.
Thus for each $j$,
$f(l) = 1$ for at least one literal $l$ in the $j$th clause,
and hence $f$ satisfies clause $\Lambda_j$.
We have shown that $f$ satisfies each clause of the 3-SAT instance,
proving that the instance is satisfiable.

\medskip
\commentpi
Conversely, suppose that the
3-SAT instance is satisfiable, so there exists a truth function $f$ satisfying it.

\commentpi
We will construct an HCLP model $H\in\calCt$ such that
$H \modelsoplus \Gamma \cup\set{\beta<\alpha}$,
and thus $H \not\modelsoplus\alpha\le\beta$,
proving that $\Gamma\not\modelsCoplust \alpha\le\beta$.

\commentpi
For $i=1,\ldots,\numpvars$,
let $S_i = \set{a_i^1, \ldots, a_i^{t-1}, \qiplus}$ if $f(p_i) = 1$,
and otherwise, let
$S_i = \set{a_i^1, \ldots, a_i^{t-1}, \qiminus}$.
We then define $H$ to be the sequence
$S_1, S_2, \ldots, S_\numpvars,  \set{c^*}$.
%
Since $\Cons(H) \ni c^*$, we have that $H \modelsoplus \beta<\alpha$.

\commentpi
By Lemma~\ref{le:GammaOne},
for all $i = 1, \ldots, \numpvars$,
$H \modelsoplus \Gamma_1^i$ and so $H \modelsoplus \Gamma_1$.
By Lemma~\ref{le:GammaTwo},
for all $i = 1, \ldots, \numpvars$,
$H \modelsoplus \vphi_i$,
so $H \modelsoplus \Gamma_2$.

\commentpi
Consider any $j \in\set{1, \ldots, s}$, and,
as above, write the $j$th clause as $l_1 \vee  l_2 \vee l_3$.
Truth assignment $f$ satisfies this clause, so
there exists $k \in \set{1, 2, 3}$ such that
$f(l_k) = 1$,
where $f$ is extended to literals in the usual way.
Then $Q(l_k)$ appears in $H$ before $c^*$,
so, by Lemma~\ref{le:GammaThree},
$H \modelsoplus \psi_j$.
Thus $H \modelsoplus \Gamma_3$.

\commentpiend
Since $\Gamma = \Gamma_1 \cup \Gamma_2 \cup \Gamma_3$,
we have shown that $H \modelsoplus \Gamma \cup \set{\beta<\alpha}$,
completing the proof.

\section{Sequence-of-Evaluations Models}
\label{sec:singleton-sequence}

In this section, we consider the case where we restrict to
HCLP models which consist of a sequence of singletons;
thus each model corresponds to a sequence of evaluations,
and generates a lexicographic order based on these.

Let $\calC$ be a set of evaluations on $\calA$.
A $\calC(1)$-model is a sequence of different elements of $\calC$.
The operation $\oplus$ plays no part,
so we can harmlessly abbreviate
ordering $\precceq_H^\oplus$ to just $\precceq_H$,
for any $\calC(1)$-model $H$,
and similarly for $\prec_H$ and $\equiv_H$.

\comment
The deduction problem for the sequence of singletons case is thus as follows.
Given $\Gamma \cup\set{\vphi}\subseteq\LA$,
is it the case that
$\Gamma \modelsCone \vphi$?
That is, is it the case that for all $\calC(1)$-models $H$ (over $\calA$),
if $H$ satisfies $\Gamma$ then $H$ satisfies $\vphi$?

\subsection{Some Basic Definitions and Results}

We write $\vphi \in \LA$
as $\alpha_\vphi < \beta_\vphi$, if $\vphi$ is strict,
or as
$\alpha_\vphi \le \beta_\vphi$,
if $\vphi$ is non-strict.
We consider a set $\Gamma\subseteq\LA$, and a set $\calC$ of evaluations on $\calA$.
For $\vphi \in \Gamma$,
define $\SuppphiC$ to be
$\set{c \in \calC \sth c(\alpha_\vphi) < c(\beta_\vphi) }$;
 define
$\OppphiC$ to be
$\set{c \in \calC \sth c(\alpha_\vphi) > c(\beta_\vphi) }$; and
 define
$\IndphiC$ to be
$\set{c \in \calC \sth c(\alpha_\vphi) = c(\beta_\vphi) }$.
We may sometimes abbreviate
$\SuppphiC$ to $\Suppphi$,
and similarly for $\OppphiC$ and $\IndphiC$.
$\Suppphi$ are the evaluations that support $\vphi$;
$\Oppphi$ are the evaluations that oppose $\vphi$.
$\Indphi$ are the other evaluations, that are indifferent regarding $\vphi$.
For a model $H$ to satisfy $\vphi$ it is necessary that
no evaluation that opposes $\vphi$ appears before
all evaluations that support $\vphi$.

\commentph
More precisely, we have the following:

\commentph
\begin{lemma}
\label{le:Hmodelsvphi}
Let $H$ be an element of
$\calCone$,
i.e., a sequence of different elements of $\calC$.
For strict $\vphi$,
$H \modelsH \vphi$ if and only if
an element of $\SuppphiC$ appears in $H$ before any
element in $\OppphiC$ appears
(and some element of $\SuppphiC$ appears in $H$).
For non-strict $\vphi$,
$H \modelsH \vphi$ if and only if
an element of $\SuppphiC$ appears in $H$ before any
element in $\OppphiC$ appears,
or no element of $\OppphiC$ appears in $H$
(i.e., $\Cons(H) \cap \OppphiC = \emp$).
\end{lemma}

\commentphstart
Let $H = (c_1, \ldots c_k)$ be a $\calC(1)$-model.
Suppose that $\vphi$ is a strict statement.
Then  $H \modelsH \vphi$, i.e. $\alpha_\vphi \prec_H \beta_\vphi$,
if and only if there exists some $i \in\set{1, \ldots, k}$ such that
 $\{ c_1, \dots c_{i-1}\} \subseteq \Indphi$ and $c_i \in \SuppphiC$,
which is if and only if
an element of $\SuppphiC$ appears in $H$ before any
element in $\OppphiC$ appears,
and some element of $\SuppphiC$ appears in $H$.

\commentphend
Now suppose that $\vphi$ is a non-strict statement.
Then $H \modelsH \vphi$, i.e. $\alpha_\vphi \precceq_H \beta_\vphi$,
if and only if
either (i)
for all $i=1, \ldots, k$, $\alpha \equiv_{c_i} \beta$;
or (ii) there exists some $i \in\set{1, \ldots, k}$ such that
  $\alpha \prec_{c_i} \beta$ and for all $j$ such that $1 \le j < i$,  $\alpha \equiv_{c_j} \beta$.
(i) holds if and only if $\Cons(H) \subseteq \Indphi$,
i.e., no element of $\SuppphiC$ or $\OppphiC$ appears in $H$.
(ii) holds if and only if
an element of $\SuppphiC$ appears in $H$ before any
element in $\OppphiC$ appears,
and some element of $\SuppphiC$ appears in $H$.
Thus, $H \modelsH \vphi$ holds if and only if
either no element in $\OppphiC$ appears in $H$
or some element of $\SuppphiC$ appears in $H$
and the first such element appears before any
element in $\OppphiC$ appears.


The following defines \emph{inconsistency bases}, which are concerned with evaluations that cannot appear in any model
satisfying the set of preference statements $\Gamma$
(see Proposition~\ref{pr:IB-basic} below).
\begin{definition}
\label{def:IB}
Let $\Gamma\subseteq\LA$, and let $\calC$ be a set of $\calA$-evaluations.
We say that $(\Gamma', C')$ is an inconsistency base for $(\Gamma, \calC)$
if $\Gamma' \subseteq\Gamma$, and
$C' \subseteq \calC$,
and
\begin{itemize}
  \item[(i)] for all $\vphi\in\Gamma'$,
  $\SuppphiC \cup \OppphiC \subseteq C'$ (and thus $\calC - C' \subseteq \IndphiC$); and
  \item[(ii)] for all $c \in C'$, there exists $\vphi \in \Gamma'$
  such that $\OppphiC \ni c$.
\end{itemize}
\end{definition}

\commentc
Note we're allowing that $C'$ be empty.
(Not allowing it, would mean we would have to define $\MIB$
for this case, which would seem to mean a more complicated version of
Proposition~\ref{pr:Gamma-allowed-MIB}.)
There is always an inconsistency base (and hence a $\MIB$) since
 $(\emp, \emp)$ is always an IB.
If $C' = \emp$ then $\Gamma'$ can only contain elements that are indifferent to all of $\calC$, and any such pair $(\Gamma', \emp)$ is an IB.
Thus, for the MIB we'll have $\Gamma^\bot$ containing
all the $\vphi$ that are indifferent to all of $\calC$.

Thus, for all $\vphi\in\Gamma'$,
$C'$ contains all evaluations that are not indifferent regarding $\vphi$,
and for all $c \in C'$ there is some element of $\Gamma'$ that is opposed by $c$.
The following result motivates the definition:

\begin{proposition}
\label{pr:IB-basic}
Let $H$ be an element of $\calCone$.
Suppose that $H \modelsH \Gamma$,
and let $(\Gamma', C')$ be an inconsistency base for $(\Gamma, \calC)$.
Then $C' \cap \Cons(H) = \emp$.
Thus no $\calCone$ model of $\Gamma$ can involve any element of $C'$.
Also, we have for any $\vphi \in \Gamma'$,
$\alpha_\vphi \equiv_H \beta_\vphi$,
so $H \not\modelsH \alpha_\vphi < \beta_\vphi$.
\end{proposition}

\commentpistartend
Let $H = (c_1, \dots c_k)$ be an element of $\calCone$ with $H \modelsH \Gamma$
	and let $(\Gamma', C')$ be an inconsistency base for $(\Gamma, \calC)$.
Suppose $H$ contains some element in $C'$
and let $c_i$ be the element in $C' \cap \Cons(H)$ with the smallest index.
By Definition~\ref{def:IB}(ii),
	there exists $\vphi \in \Gamma'$ such that $\OppphiC \ni c_i$.
Furthermore, since $c_j \notin C'$ for all $1\leq j < i$,
Definition~\ref{def:IB}(i) implies $c_j \in \IndphiC$.
	But then, an evaluation that opposes $\vphi$ appears before all evaluations that support $\vphi$.
	By Lemma~\ref{le:Hmodelsvphi}, this is a contradiction to $H \modelsH \Gamma$;
hence we must have $C' \cap \Cons(H) = \emp$.
	Also, for all $\vphi\in\Gamma'$, $\SuppphiC \cup \OppphiC \subseteq C'$ and thus $ \Cons(H) \subseteq \calC - C' \subseteq \IndphiC$.
	Hence, for any $\vphi \in \Gamma'$, $\alpha_\vphi \equiv_H \beta_\vphi$.

Thus $H$ does not strictly satisfy any element of $\Gamma'$.

\commentc
Do I define ``strictly satisfy"?
Maybe this is OK.

\ssk
For inconsistency bases $(\Gamma_1, C_1)$ and $(\Gamma_2, C_2)$
for $(\Gamma, \calC)$,
define $(\Gamma_1, C_1) \cup (\Gamma_2, C_2)$
to be $(\Gamma_1 \cup \Gamma_2, C_1 \cup C_2)$.
It is easy to show that if
$(\Gamma_1, C_1)$ and $(\Gamma_2, C_2)$
are both inconsistency bases for $(\Gamma, \calC)$
then
 $(\Gamma_1, C_1) \cup (\Gamma_2, C_2)$ is one also.

\commentpi
\begin{lemma}
\label{le:IB-union}
If $(\Gamma_1, C_1)$ and $(\Gamma_2, C_2)$
are both inconsistency bases for $(\Gamma, \calC)$,
then $(\Gamma_1, C_1) \cup (\Gamma_2, C_2)$
is an inconsistency base for $(\Gamma, \calC)$.
\end{lemma}

\commentpistartend
	Let $(\Gamma_1, C_1)$ and $(\Gamma_2, C_2)$ be two inconsistency bases for $(\Gamma, \calC)$.
	Then by Definition~\ref{def:IB} (i),
	for all $\vphi\in\Gamma_1$, $\SuppphiC \cup \OppphiC \subseteq C_1$; and
  	for all $\vphi\in\Gamma_2$, $\SuppphiC \cup \OppphiC \subseteq C_2$.
  	Thus, for all $\vphi\in\Gamma_1 \cup \Gamma_2$, $\SuppphiC \cup \OppphiC \subseteq C_1 \cup C_2$.
  	Hence, $(\Gamma_1, C_1) \cup (\Gamma_2, C_2)$ satisfies condition (i) of the definition of inconsistency bases.
  	By Definition~\ref{def:IB} (ii),
  	for all $c \in C_1$, there exists $\vphi \in \Gamma_1$ such that $\OppphiC \ni c$; and
  	for all $c \in C_2$, there exists $\vphi \in \Gamma_2$ such that $\OppphiC \ni c$.
  	Thus, for all $c \in C_1 \cup C_2$, there exists $\vphi \in \Gamma_1 \cup \Gamma_2$ such that $\OppphiC \ni c$.
  	And hence, $(\Gamma_1, C_1) \cup (\Gamma_2, C_2)$ satisfies condition (ii) of the definition of inconsistency bases.

Define
$\MIB(\Gamma, \calC)$, the
Maximal Inconsistency Base for $(\Gamma, \calC)$,  to be the union of all inconsistency bases
for  $(\Gamma, \calC)$,
i.e.,
$\bigcup\set{(\Gamma_1, C_1) \in \calI}$,
where $\calI$ is the set of inconsistency bases for $(\Gamma, \calC)$.
The next result follows.

\begin{proposition}
\label{pr:MIB}
$\MIB(\Gamma, \calC)$ always exists, and is an inconsistency base for $(\Gamma, \calC)$,
which is maximal in the following sense:
if $(\Gamma_1, C_1)$ is an inconsistency base for $(\Gamma, \calC)$
then $\Gamma_1 \subseteq \Gamma^\bot$ and $C_1 \subseteq C^\bot$,
where $\MIB(\Gamma, \calC) = (\Gamma^\bot, C^\bot)$.
\end{proposition}

\commentpistartend
	The trivial tuple $(\emptyset,\emptyset)$ is always an inconsistency base.
	Hence, since $\MIB(\Gamma, \calC)$ is defined to be the union of all inconsistency bases, it must always exist.
	By Lemma~\ref{le:IB-union}, the union of two inconsistency bases is an inconsistency base.
	Consequently, $\MIB(\Gamma, \calC)$ is an inconsistency base.
	Furthermore, $\MIB(\Gamma, \calC) = (\Gamma^\bot, C^\bot)$ is defined to be $(\bigcup_{(C',\Gamma') \in \calI} C', \bigcup_{(C',\Gamma') \in \calI} \Gamma')$,
	where $\calI$ is the set of inconsistency bases for $(\Gamma, \calC)$.
	Thus, if $(\Gamma_1, C_1)$ is an inconsistency base for $(\Gamma, \calC)$,
then $\Gamma_1 \subseteq \Gamma^\bot$ and $C_1 \subseteq C^\bot$.

Note, that $\MIB(\Gamma, \calC)$ is the unique Maximal Inconsistency Base.
It can also be easily shown that
if $\Gamma$ is $\calCone$-consistent then
$\Gamma^\bot$ contains no strict elements. Theorem~\ref{th:algorithm} below implies the converse of this result.

\commentpi
\begin{proposition}
Suppose that $\Gamma$ is $\calCone$-consistent,
i.e., there exists a $\calCone$ model of $\Gamma$.
Then for any Inconsistency Base $(\Gamma', C')$ of $(\Gamma, \calC)$,
$\Gamma' \cap \Gamma^< = \emp$,
(where $\Gamma^<$ is the set of strict elements of $\Gamma$).
In particular, if
$(\Gamma', C') = \MIB(\Gamma, \calC)$
then $\Gamma' \cap \Gamma^< = \emp$.
\end{proposition}

\commentpistartend
By Proposition~\ref{pr:IB-basic},
for every inconsistency base $(\Gamma', C')$ and $H \in \calCone$ with $H \modelsH \Gamma$,
we have for any $\vphi \in \Gamma'$,
$H \not\modelsH \alpha_\vphi < \beta_\vphi$.
Thus, if $\Gamma$ is $\calCone$-consistent,
i.e., there exists some $H \in \calCone$ with $H \modelsH \Gamma$,
then $\Gamma'$ contains no strict statements.

\subsection{A Polynomial Algorithm for Consistency and Deduction}
\label{subsec:Algorithm}

Throughout this section we consider a
set $\Gamma \subseteq\LA$ of input preference statements,
and a set $\calC$ of $\calA$-evaluations.


Define
$\Opp_\Gamma(c)$ (abbreviated to $\Opp(c)$)
to be the set of elements opposed by $c$,
i.e., $\vphi\in\Gamma$ such that $c(\alpha_\vphi) > c(\beta_\vphi)$,
and define
$\Supp_\Gamma(c)$ (usually abbreviated to $\Supp(c)$) to be the set of elements $\vphi$ of $\Gamma$ supported by $c$,
(i.e.,
$c(\alpha_\vphi) < c(\beta_\vphi)$).
Also, for sequence of evaluations $(c_1,\ldots, c_k)$,
we define $\Supp(c_1,\ldots, c_k)$ to be $\bigcup_{i =1}^k \Supp(c_i)$.

\commenta
Previous version of last paragraph:
For $C' \subseteq \calC$, define
$\Supp_\Gamma(C')$ (usually abbreviated to $\Supp(C')$) to be the elements of $\Gamma$ that are supported by some
element of $C'$.
Thus,
$\Supp(C') = \bigcup_{c\in C'} \Supp(c)$.
Also, for sequence of evaluations $(c_1,\ldots, c_k)$,
we define $\Supp(c_1,\ldots, c_k)$ to be $\Supp(\set{c_1,\ldots, c_k})$.

The idea behind the algorithm is as follows: suppose that we have picked a sequence of elements of $\calC$,
$C'$ being the set picked so far.
We need to choose next an evaluation $c$
such that,
if $c$ opposes some $\vphi$ in $\Gamma$,
 then $\vphi$ is supported by some evaluation in $C'$
(or else the generated sequence will not satisfy $\vphi$).

\commentb
We have $H \modelsH \Gamma$ implies $H \modelsH \Gammadestrict$,
so $H \not\modelsH \Gammadestrict$
implies $H \not\modelsH \Gamma$.

\subsubsection*{The Algorithm}

\newcommand{\ntab}{\quad}
\newcommand\Conscheck{\textit{Cons-check}}

$H$ is initialised as the empty list $()$ of evaluations.
$H \leftarrow H + c$ means that evaluation $c$ is added to the end of $H$.
\msk

\noindent
Function $\Conscheck(\Gamma, \calC)$
\ssk

\noindent
$H\leftarrow ()$

\noindent
\textbf{for} $k =1, \ldots, |\calC|$ do

\noindent
\ntab\textbf{if}
$\exists$ $c \in \calC - \Cons(H)$ such that $\Opp(c) \subseteq \Supp(H)$

\noindent
\ntab\ntab\textbf{then} choose some such $c$; \  $H \leftarrow H + c$

\noindent
\ntab\ntab\textbf{else} \textbf{stop}

\noindent
\ntab\textbf{end} \textbf{for}

\noindent
\textbf{return} $H$

\msk
The algorithm
involves often non-unique choices. However, if we wish, the choosing of $c$ can be done based on an ordering
$c_1, \ldots, c_m$ of $\calC$, where, if there exists more than one $c \in \calC - \Cons(H)$ such that $\Opp(c) \subseteq \Supp(H)$,
we choose the element $c_i$ with smallest index $i$.
The algorithm then becomes deterministic, with a unique result following from the given inputs.

A straight-forward implementation runs in $O(|\Gamma| |\calC|^2)$ time;
however, a more careful implementation runs in $O(|\Gamma| |\calC|)$ time.

We describe this implementation in the following.
Let $H_k$ be the HCLP model after the $k$-th iteration of the for-loop.
In every iteration of the for-loop,
we update sets $O_k^{\Delta}(c) = \Opp(c) - \Supp(H_k)$
and $S_k^{\Delta}(c) = \Supp(c) - \Supp(H_k)$
for all $c \in \calC - \Cons(H_k)$.
This costs us $O(|\calC - \Cons(H_k)|\times|\Supp(H_{k}) \setminus \Supp(H_{k-1})|)
 = O(|\calC - \Cons(H_k)|\times|S_{k-1}^{\Delta}(c_k)|)$
 more time for every iteration $k$ in which we add evaluation $c_k$ to $H_{k-1}$.
 However, the choice of the next evaluation $c_k$ be performed in constant time by marking evaluations $c$ with $O_{k-1}^{\Delta}(c) = \emptyset$.
 Suppose the algorithm stops after $1 \leq l \leq |\calC|$ iterations.
 Since all $S_{k-1}^{\Delta}(c_k)$ are disjoint,
 $\sum_{k=1}^{l} |S_{k-1}^{\Delta}(c_k)| = |\Supp(H_l)| \leq |\Gamma|$.
 Altogether, the running time is $O(\sum_{k=1}^{l} |\calC - \Cons(H_k)|\times|S_{k-1}^{\Delta}(c_k)|)
  \le O(|\calC|\times\sum_{k=1}^{l} |S_{k-1}^{\Delta}(c_k)|)$,
  and thus the running time is
  $O(|\calC|\times|\Gamma|)$.

\subsubsection*{Properties of the Algorithm}

Any $\vphi\in\Gamma$ which is opposed by some evaluation $c$ in $H$
is supported by some earlier evaluation in $H$.
Consider any $\vphi$ in $\Supp(H)$.
Let $c_j$ be the earliest evaluation in $H$ that supports $\vphi$.
None of the earlier evaluations than $c_j$ oppose $\vphi$,
and thus $H$ strictly satisfies $\vphi$.
A similar argument shows
that
$H$ satisfies  $\Gammadestrict$, defined to be
$\set{\alpha_\vphi \le \beta_\vphi \sth \vphi\in\Gamma}$,
i.e., $\Gamma$ where the strict statements are replaced
by corresponding non-strict statements.

\comment
$H$ satisfies $\alpha_\vphi \le \beta_\vphi$ if and only if
the earliest evaluation in $H$
which is not indifferent to $\vphi$
supports (rather than opposes) $\vphi$.

The algorithm will always generate an HCLP model satisfying $\Gamma$ if $\Gamma$ is consistent.
It can also be used for computing the Maximal Inconsistency Base.
The following result sums up some properties related to the algorithm.

\begin{theorem}
\label{th:algorithm}
Let $H$ be a sequence returned by the algorithm
with inputs $\Gamma$ and $\calC$, and
write $\MIB(\Gamma, \calC)$  as $(\Gamma^\bot, C^\bot)$.
Then  $C^\bot = \calC - \Cons(H)$
(i.e., the evaluations that don't appear in $H$),
and $\Gamma^\bot = \Gamma - \Supp(H)$.
We have that $H \modelsH \Gammadestrict$.
Also, $\Gamma$ is $\calCone$-consistent if and only
if $\Supp(H)$ contains all the strict elements of $\Gamma$,
which is if and only if $\Gamma^\bot \cap \Lstrict = \emp$.
If $\Gamma$ is $\calCone$-consistent then
$H\modelsH \Gamma$.
\end{theorem}

\commentphstart
	Because of the if-condition $\Opp(c) \subseteq \Supp(H)$, we never add an evaluation $c$ to $H$ throughout the algorithm that opposes some not yet supported preference statement.
In particular, because of Lemma~\ref{le:Hmodelsvphi}, this implies that
the returned sequence $H$ satisfies every element of  $\Gammadestrict$.

\commentph
	In order to prove the remaining statements, first observe the following.
	The algorithm terminates only if for all $c \in \calC - \Cons(H)$, $\Opp(c) \nsubseteq \Supp(H)$, i.e., $\Opp(c) - \Supp(H) \neq \emptyset$.
	Thus, for all $c \in \calC - \Cons(H)$, there exists $\vphi \in \Gamma - \Supp(H)$ such that $c \in \OppphiC$.
	Furthermore, since $H$ satisfies $\Gammadestrict$, we have that
	for all $\vphi \in \Gamma - \Supp(H)$, $\Cons(H) \subseteq \IndphiC$
(and thus $\SuppphiC \cup \OppphiC \subseteq \calC - \Cons(H)$).
	Hence, $(\calC - \Cons(H), \Gamma - \Supp(H))$ satisfies (i) and (ii) of Definition~\ref{def:IB} and thus is an inconsistency base.

\commentph
By Proposition~\ref{pr:MIB}, $C^\bot \supseteq \calC - \Cons(H)$ and $\Gamma^\bot \supseteq \Gamma - \Supp(H)$.
We will show that $C^\bot \subseteq \calC - \Cons(H)$ and $\Gamma^\bot \subseteq \Gamma - \Supp(H)$,
and thus $C^\bot = \calC - \Cons(H)$ and $\Gamma^\bot = \Gamma - \Supp(H)$.
We have that $\MIB(\Gammadestrict, \calC) = \MIB(\Gamma, \calC) = (\Gamma^\bot, C^\bot)$,
since the definition of $\MIB$ does not depend on whether statements are strict or not.
Proposition~\ref{pr:IB-basic} implies that $C^\bot \cap \Cons(H) = \emp$,
and thus $C^\bot \subseteq \calC - \Cons(H)$.
	Next, we show  $\Gamma^\bot \subseteq \Gamma - \Supp(H)$ and hence $\Gamma^\bot = \Gamma - \Supp(H)$.
	By Definition~\ref{def:IB} (i), $\Cons(H) = \calC - C^\bot \subseteq \IndphiC$ for all $\vphi \in \Gamma^\bot$.
	Thus, $\Gamma^\bot \cap \Supp(H) = \emptyset$, i.e., $\Gamma^\bot \subseteq \Gamma - \Supp(H)$.

\commentphend
	Now, $\Supp(H)$ contains all the strict elements of $\Gamma$ if and only if $(\Gamma - \Supp(H)) \cap \Lstrict = \Gamma^\bot \cap \Lstrict = \emp$.
	Proposition~\ref{pr:IB-basic} implies, if there exists $H \in \calCone$ with $H \modelsH \Gamma$, then $\Gamma^\bot$ can contain no strict statements, i.e., $\Gamma^\bot \cap \Lstrict = \emp$.
	Now suppose, $\Gamma^\bot \cap \Lstrict = \emp$.
	Then all strict statements are contained in $ \Gamma - \Gamma^\bot = \Supp(H)$.
	Thus, all strict statements are satisfied by $H$ and since $H$ satisfies $\Gammadestrict$, $\Gamma$ is $\calCone$-consistent.
	Hence, $\Gamma$ is $\calCone$-consistent if and only if $\Supp(H)$ contains all the strict elements of $\Gamma$.
	Also, if $\Gamma$ is $\calCone$-consistent, then $H\modelsH \Gamma$.

The algorithm therefore determines $\calCone$-consistency,
 and hence $\calCone$-deduction (because of Proposition~\ref{pr:basic-cons-deduction}), in polynomial time,
and also generates the Maximal Inconsistency Base.
For the case when $\Gamma$ is not $\calCone$-consistent,
the output $H$ of
the algorithm is a model which, in a sense, comes closest to satisfying $\Gamma$:
it always satisfies
$\Gammadestrict$, the non-strict version of $\Gamma$,
and if any model $H' \in\calCone$ satisfies $\Gammadestrict$ and any element
$\vphi$ of $\Gamma$,
then $H$ also satisfies $\vphi$.
For, if $H\not\modelsH \vphi$, then $\vphi$ is strict
(since $H$ satisfies all the non-strict elements of $\Gamma$, because $H \modelsH \Gammadestrict$).
 Also, $\vphi\notin\Supp(H)$ (since $H \modelsH\Supp(H)$),
and thus, $\vphi \in \Gamma^\bot$, where $(\Gamma^\bot, C^\bot) = \MIB(\Gamma, \calC) = \MIB(\Gammadestrict, \calC)$.
Proposition~\ref{pr:IB-basic} implies that $H'\not \modelsH \vphi$, since $\vphi$ is strict.
Proposition~\ref{pr:IB-basic} also implies that $\Cons(H') \cap C^\bot = \emp$,
and thus $\Cons(H') \subseteq\Cons(H) = \calC - C^\bot$, using Theorem~\ref{th:algorithm}.
We have proved the following result.

\begin{lemma}
\label{le:alg-max-model}
Let $H$ be a sequence returned by the algorithm
with inputs $\Gamma$ and $\calC$,
and suppose that $H' \in\calCone$ is such that $H' \modelsH \Gammadestrict$.
Then, for all $\vphi \in \Gamma$,
if $H' \modelsH \vphi$ then $H \modelsH \vphi$.
Also, $\Cons(H') \subseteq \Cons(H) = \calC - C^\bot$.
\end{lemma}


These properties suggest the following way of reasoning with inconsistent $\Gamma$.
Let us define $\Gamma'$ to be equal to $(\Gamma - \Gamma^\bot) \cup \Gammadestrict$.
By Theorem~\ref{th:algorithm}, this is equal to $\Supp(H) \cup \Gammadestrict$,
where $H$ is a model generated by the algorithm,
enabling easy computation of $\Gamma'$.
$\Gamma'$ is consistent, since it is satisfied by $H$.
We might then (re-)define the (non-monotonic) deductions from inconsistent $\Gamma$
to be the deductions from $\Gamma'$.

\commentc
This last/middle part follows from results in longer document:
by construction of the algorithm, and the definition of maximal $\Gamma$-allowed sequence,
the output of the algorithm is a maximal $\Gamma$-allowed sequence;
Proposition~\ref{pr:max-Gamma-allowed-models} then implies this property.

\newcommand{\calCprimeone}{C'(1)}

\commenta
Check notation for $\modelsCprimeone$ etc
Note that I'm using $\modelsCprimeone$ in two different places,
? and slightly different ways. Does this matter?

\subsection{Strong Consistency}

In the set of models $\calCone$, we allow models involving any subset of $\calC$, the set of evaluations.
We could alternatively consider a semantics where we
 only allow models $H$ that involve all elements of $\calC$,
i.e., with $\Cons(H) = \calC$.

Let $\calConestar$ be the set of elements $H$ of $\calCone$
with $\Cons(H) = \calC$.
 $\Gamma$ is defined to be \emph{strongly $\calCone$-consistent} if and only if
there exists a model $H\in\calConestar$ such that
$H \modelsH \Gamma$.
Let $\MIB(\Gamma, \calC) = (\Gamma^\bot, C^\bot)$.
Proposition~\ref{pr:IB-basic}
implies that, if $\Gamma$ is strongly $\calCone$-consistent
then $C^\bot$ is empty, and
$\Gamma^\bot$ consists of all the elements of $\Gamma$
that are indifferent to all of $\calC$,
i.e., the set of $\vphi\in\Gamma$ such that
$c(\alpha_\vphi) = c(\beta_\vphi)$ for all $c \in \calC$.

\newcommand{\calMCmax}{\calM_{\calCone}^{\mathrm{max}}}
\newcommand{\modelsCmax}{\models_{\calCone}^{\mathrm{max}}}

We write $\Gamma \modelsConestar \vphi$ if
$H \modelsH \vphi$ holds
for every $H \in \calConestar$ such that
$H \modelsH \Gamma$.
The next result shows that the non-strict
$\modelsConestar$ inferences are the same
as the non-strict $\modelsCone$ inferences,
and that (in contrast to the case of $\modelsCone$),
the strict $\modelsConestar$ inferences
 almost correspond with the non-strict ones.
The result also implies that the algorithm in Section~\ref{subsec:Algorithm} can be used to
efficiently determine the $\modelsConestar$ inferences.

\commentph
This form of deduction can be expressed in terms of strong consistency, as the following result shows.

\commentph
\begin{lemma}
\label{le:Strong-cons-vs-deduction}
If $\Gamma$ is  strongly $\calCone$-consistent,
then
$\Gamma \modelsConestar \vphi$ holds if and only if
$\Gamma \cup\set{\neg\vphi}$ is not strongly $\calCone$-consistent.
\end{lemma}

\commentphstart
Assume that $\Gamma$ is  strongly $\calCone$-consistent.
First suppose that
$\Gamma \cup\set{\neg\vphi}$ is strongly consistent.
Then there exists
$H\in\calCone$ such that $H\modelsH \Gamma\cup\set{\neg\vphi}$
and $\Cons(H) = \calC$.
Thus $H\modelsH \Gamma$ and $H\not\modelsH\vphi$
(using Lemma~\ref{le:basic-negation}),
showing that $\Gamma \not\modelsConestar \vphi$.

\commentphend
Now suppose that $\Gamma \not\modelsConestar \vphi$.
Then there exists
$H \in \calCone$ such that
$H \modelsH \Gamma$ and $\Cons(H) = \calC$
and $H\not\modelsH\vphi$.
Then $H\modelsH \cup\set{\neg\vphi}$
(again using Lemma~\ref{le:basic-negation}),
so $\Gamma \cup\set{\neg\vphi}$ is strongly $\calCone$-consistent.

\commenta
**
Some lemmas could be added and proved for the proof of Case(i) below.
The next sentence is new.

We write
$\Gamma \modelsConestar \alpha \equiv \beta$
as an abbreviation of the conjunction of
$\Gamma \modelsConestar \alpha \le \beta$
and
$\Gamma \modelsConestar \beta \le \alpha$.

\begin{proposition}
\label{pr:strong-cons-properties}
Let $\MIB(\Gamma, \calC) = (\Gamma^\bot, C^\bot)$.
$\Gamma$ is strongly $\calCone$-consistent if and only if
$C^\bot = \emp$ and $\Gamma \cap \Lstrict \subseteq \Supp(\calC)$.\footnotemark

Suppose that $\Gamma$ is strongly $\calCone$-consistent.
Then,
\begin{itemize}
  \item[(i)] $\Gamma \modelsCone \alpha \le \beta$
$\iff$ $\Gamma \modelsConestar \alpha \le \beta$;
  \item[(ii)] $\Gamma \modelsConestar \alpha \equiv \beta$
  if and only if $\alpha$ and $\beta$ agree on all of $\calC$,
  i.e., for all $c \in\calC$, $c(\alpha) = c(\beta)$;
  \item[(iii)] $\Gamma \modelsConestar \alpha < \beta$
  if and only if $\Gamma \modelsCone \alpha \le \beta$ and
  $\alpha$ and $\beta$ differ on some element of $\calC$,
  i.e., there exists $c\in\calC$ such that $c(\alpha) \not= c(\beta)$.
\end{itemize}
\end{proposition}

\footnotetext{Note, that at this point the original paper states:
``$\Gamma$ is strongly $\calCone$-consistent if and only if $C^\bot = \emp$''.
However, this is not quite correct,
because of the (somewhat pathological) case where
there exists an element $\alpha < \beta$ in $\Gamma$ with
all evaluations agreeing on $\alpha$ and $\beta$,
i.e., for all $c \in \calC$, $c(\alpha) = c(\beta)$;
such a $\vphi$ is inconsistent on its own.
(In fact, if $C^\bot = \emp$ then
$\Gamma$ is not strongly $\calCone$-consistent if and only there exists some
such self-inconsistent element of $\Gamma$.)
See e.g., the following example
with alternatives $\calA = \set{\alpha, \beta, \gamma}$, preference statements $\Gamma = \set{\alpha < \beta, \beta < \gamma}$ and with evaluations $\calC = \set{c_1, c_2}$, defined as follows.\\
 $c_1(\alpha) = 0$; \  $c_1(\beta) = 2$; \  $c_1(\gamma) = 2$;\\
 $c_2(\alpha) = 2$; \  $c_2(\beta) = 1$; \  $c_2(\gamma) = 1$.\\
  Then $C^\bot = \emp$, but $\Gamma$ is not strongly $\calCone$-consistent since the statement $\beta < \gamma$ is not satisfied by any model in $\calCone$.}

\newcommand{\Suppab}{\textit{Supp}^{\alpha \le \beta}}
\newcommand{\Oppab}{\textit{Opp}^{\alpha \le \beta}}
\newcommand{\Ind}{\textit{Ind}}

 \commentphstart
First, suppose that $\Gamma$ is strongly $\calCone$-consistent.
Then there exists
$H'\in\calCone$ such that $H'\modelsH \Gamma$
and $\Cons(H') = \calC$.
Let $H$ be a model generated by the algorithm.
We will show that $H\modelsH \Gamma$
and $\Cons(H) = \calC$.
Lemma~\ref{le:alg-max-model} implies that
$H$ satisfies every element of $\Gamma$, i.e., $H \modelsH \Gamma$, since $H'\modelsH \Gamma$;
the lemma also implies that
$\Cons(H') \subseteq \Cons(H)\subseteq\calC$,
so $\Cons(H) = \calC$.
Proposition~\ref{pr:IB-basic} implies that $C^\bot = \emp$.
Theorem~\ref{th:algorithm} implies that
$\Supp(\calC)$ contains all the strict elements of $\Gamma$, i.e.,
 $\Gamma \cap \Lstrict \subseteq \Supp(\calC)$.

\commentph
Conversely, suppose that
$C^\bot = \emp$ and $\Gamma \cap \Lstrict \subseteq \Supp(\calC)$.
Let $H$ be a model generated by the algorithm.
Theorem~\ref{th:algorithm} implies that
$\Cons(H) = \calC - C^\bot$, i.e., $\Cons(H) = \calC$,
and also $H \modelsH \Gammadestrict$.
Since $H \modelsH \Supp(H)$,
we have $H \modelsH \Gamma \cap \Lstrict$
since $\Supp(H) = \Supp(\calC)$.
Thus $H \modelsH \Gamma$,
since $\Gamma \subseteq \Gammadestrict \cup (\Gamma \cap \Lstrict)$.
This implies that $\Gamma$ is strongly $\calCone$-consistent.

\commentph
	Now suppose that $\Gamma$ is strongly $\calCone$-consistent. We consider (i)--(iii) in turn.
	\begin{itemize}
 	 \item[(i)] If $\Gamma \modelsCone \alpha \le \beta$,
 	 		then $\Gamma \modelsConestar \alpha \le \beta$, because $\calConestar \subseteq \calCone$.
 	 		Conversely, suppose that $\Gamma \not\modelsCone \alpha \le \beta$.
            Then there exists $H' \in \calCone$ with $H' \modelsH \Gamma\cup\set{\beta < \alpha}$.
           Suppose that $H$ is generated by the algorithm applied to $\calC$ and $\Gamma$,
           where we break ties using an order extending (i.e., starting with) $H'$.
           It can be seen that $H$ extends $H'$, which implies that $H \modelsH \beta < \alpha$.
           The argument at the beginning of the proof implies that $H\modelsH \Gamma$ and $\Cons(H) = \calC$,
           since $\Gamma$ is strongly $\calCone$-consistent.
           We have $H\modelsH \Gamma$ and $H\not\modelsH \alpha \le \beta$, and $H\in \calConestar$,
           and hence, $\Gamma \not\modelsConestar \alpha \le \beta$.
 	 \item[(ii)] If for all $c \in\calC$, $c(\alpha) = c(\beta)$, then clearly $\Gamma \modelsConestar \alpha \equiv \beta$.
 	 		Now, suppose, $\Gamma \modelsConestar \alpha \equiv \beta$.
 	 		Since $\Gamma$ is strongly $\calCone$-consistent, there exists a model $H$ in $\calConestar$ with
            $H\modelsH \Gamma$ and thus $H \modelsH \alpha \equiv \beta$.
 	 		Therefore, for all $c \in\calC$, $c(\alpha) = c(\beta)$, since $\Cons(H) = \calC$.
   	\item[(iii)] First suppose that $\Gamma \modelsConestar \alpha < \beta$.
   This implies that $\Gamma \modelsConestar \alpha \le \beta$.
   Part (i) implies that $\Gamma \modelsCone \alpha \le \beta$.
   If it were not the case that $\alpha$ and $\beta$ differ on some element of $\calC$
   then clearly for any $H\in\calCone$, $H \not\modelsH \alpha < \beta$,
   contradicting $\Gamma \modelsConestar \alpha < \beta$.
   Conversely, suppose that  $\Gamma \modelsCone \alpha \le \beta$ and
  $\alpha$ and $\beta$ differ on some element of $\calC$.
  Consider any $H \in \calConestar$ such that $H \modelsH \Gamma$; we need to show that $H \modelsH \alpha < \beta$.
  Since, $\Gamma \modelsCone \alpha \le \beta$, we have $H \modelsH \alpha \le \beta$, i.e., $\alpha \precceq_H \beta$.
  Because $H$ involves every element of $\calC$ (since $\Cons(H) = \calC$),
  and $\alpha$ and $\beta$ differ on some element of $\calC$, we have $\alpha \prec_H \beta$,
  and thus  $H \modelsH \alpha < \beta$, as required.
	\end{itemize}


We obtain a similar (and generalisation of this) result,
if we consider only the maximal models in $\calCone$ that satisfy $\Gamma$
(these all have the same cardinality: $|\calC| - |C^\bot|$).

The next result shows that $\modelsCone$ inference is not affected if one removes the evaluations in the MIB.

\begin{proposition}
\label{pr:reducing-C}
Suppose that $\Gamma$ is $\calCone$-consistent,
let $\MIB(\Gamma, \calC) = (\Gamma^\bot, C^\bot)$,
and let $C' = \calC - C^\bot$.
Then $\Gamma$ is strongly $\calCprimeone$-consistent,
and
$\Gamma \modelsCone \vphi$ if and only if
$\Gamma \modelsCprimeone \vphi$.
\end{proposition}

\commentphstartend By Theorem~\ref{th:algorithm}, any output of the algorithm is in $C'(1^*)$ and satisfies $\Gamma$.
	Thus $\Gamma$ is strongly $\calCprimeone$-consistent.
	Let $\calH' = \{H \in \calCprimeone \; : \; H \modelsH \Gamma\}$
and $\calH = \{H \in \calCone \; : \; H \modelsH \Gamma\}$.
	Then $\calH' \subseteq \calH$, because $\calCprimeone \subseteq \calCone$.
	By Proposition~\ref{pr:IB-basic}, for every $H \in \calH$, we have $\Cons(H) \cap C^\bot = \emp$, and hence $H \in \calH'$.
	Thus $\calH' = \calH$ and $\Gamma \modelsCone \vphi$ if and only if
$\Gamma \modelsCprimeone~\vphi$.

\subsection{Orderings on evaluations}

The preference logic defined here is closely related to a logic based on disjunctive ordering statements.
Given set of evaluations $\calC$,
we consider the set of statements $\calO_\calC$ of the form
$C_1 < C_2$,
and of $C_1 \le C_2$,  where $C_1$ and $C_2$ are disjoint subsets
of $\calC$.

We say that $H \models C_1 < C_2$ if
some evaluation in $C_1$ appears in $H$ before every element of $C_2$,
that is,
there exists some element of $C_1$ in $H$
(i.e., $C_1 \cap \Cons(H) \not=\emp$)
and the earliest element of $C_1 \cup C_2$ to appear in $H$
is in $C_1$.

We say that $H \models C_1 \le C_2$
if either $H \models C_1 < C_2$
or no element of $C_1$ or $C_2$ appears in $H$:
 $(C_1 \cup C_2) \cap \Cons(H) =\emp$.
Then we have that

$H \models \alpha_\vphi < \beta_\vphi$  $\iff$
$H \models \SuppphiC < \OppphiC$,

and
$H \models \alpha_\vphi \le \beta_\vphi$  $\iff$
$H \models \SuppphiC \le \OppphiC$.

\noindent
This shows that the language $\calO_\calC$ can express anything that can be expressed in $\LA$.
It can be shown, conversely, that for
any statement $\tau$ in $\calO_\calC$, one can
define $\alpha_\vphi$ and $\beta_\vphi$, and the values of elements of $\calC$ on these, such that
for all $H \in \calCone$,
$H\modelsH \tau$ if and only if $H \modelsH \vphi$.
For instance, if $\tau$ is the statement $C_1 < C_2$,
we can define $c(\alpha_\vphi) = 1$ for all $c \in C_2$,
and $c(\alpha_\vphi) = 0$ for $c\in\calC - C_2$;
and define $c(\beta_\vphi) = 1$ for all $c \in C_1$,
and $c(\beta_\vphi) = 0$ for $c\in\calC - C_1$.

\commenta
Check notation!

The algorithm adapts in the obvious way to the case where we have $\Gamma$ consisting of (or including) elements in $\calO_\calC$. When viewed in this way, the algorithm can be seen as a simple extension of a topological sort algorithm;
 the standard case corresponds to when the ordering statements
 only consist of singleton sets.

\commenta
I'm now cutting this statement:
Inconsistency Bases can be seen to be a generalised form of cycle, when this kind of the disjunctive ordering statements are allowed.

\commenta
Citation for topological sort?

\comment
$\Supp(H)$ then corresponds to the set of ordering statements that have already been shown to be (strictly) satisfied.
(They are the ordering statements such that some $c$ in $H$ appears in the left-hand-side.
Note that the algorithm never selects an evaluation $c\in\calC$
to place in $H$
which appears in the right-hand-side of an ordering statement, unless an earlier element of $H$ appears in the left-hand-side of the ordering statement (and thus the statement has already been satisfied).)
Choose evaluation $c$ that does not appear in the right hand side of any ordering statement which is yet to be satisfied.
When viewed in this way, the algorithm can be seen as a simple extension of a topological sort algorithm;
 the standard case corresponds to when the ordering statements
 only consist of singleton sets.

\commenta
? - Should I talk about criteria rather than evaluations??

\section{Discussion and Conclusions}
\label{sec:Conclusions}


We defined a class of relatively simple preference logics based on hierarchical models.
These generate an adventurous form of inference,
which can be helpful if there is only relatively sparse input preference information.
We showed that the complexity of
preference deduction is \conp-complete in general,
and polynomial for the case where the criteria are assumed to be totally ordered
(the sequence-of-evaluations case, Section~\ref{sec:singleton-sequence}).

The latter logic has strong connections with the preference inference formalism described in \cite{WilsonECAI14}.
To clarify the connection, for each evaluation $c\in\calC$ we can generate a variable $X_c$,
and let $V$ be the set of these variables.
For each alternative $\alpha\in\calA$ we generate a complete assignment $\alpha^*$
on the variables $V$ (i.e., an outcome as defined in \cite{WilsonECAI14})
by $\alpha^*(X_c) = c(\alpha)$ for each $X_c \in V$.
Note that values of $\alpha^*(X_c)$ are non-negative numbers,
and thus  have a fixed ordering, with zero being the best value.
A preference statement $\alpha \le \beta$ in $\Lnonstrict$ then corresponds with
a basic preference formula $\alpha^* \ge \beta^*$ in \cite{WilsonECAI14}.
Each model $H \in \calCone$
corresponds to a sequence of evaluations, and thus
has an associated sequence of variables; this sequence together with the fixed value orderings,
generates a lexicographic model as defined in \cite{WilsonECAI14}.

In contrast with the lexicographic inference system in \cite{WilsonECAI14},
the logic developed in this paper allows strict (as well as non-strict) preference statements,
and allows more than one variable at the same level.
However, the lexicographic inference logic from \cite{WilsonECAI14} does not assume a fixed value ordering
(which, translated into the current formalism,
corresponds to not assuming that the values of the evaluation function are known);
it also allows a richer language of preference statements, where a statement can be a compact representation for a (possibly exponentially large) set of
basic preference statements of the form $\alpha \le \beta$.
Many of the results of Section~\ref{sec:singleton-sequence} immediately extend to richer preference languages (by replacing a preference statement by a corresponding set of basic preference statements).
In future work we will determine under what circumstances deduction remains polynomial when extending the language, and when removing the assumption that the evaluation functions are known.

The \conp-hardness result for the general case (and for the $\modelsCoplust$
systems with $\cardup \ge 2$) is notable and perhaps surprising, since these preference logics are relatively simple ones.
The result obviously extends to more general systems.
The preference inference system described in \cite{WilsonIJCAI09} is based on much more complex forms of lexicographic models, allowing conditional dependencies, as well as having local orderings on sets of variables (with bounded cardinality).
Theorem~\ref{th:main-complexity-theorem} implies that
the (polynomial) deduction system in \cite{WilsonIJCAI09} is not more general than the system described here (assuming $\ppp\not=\np$).
It also implies that if one were to extend the system from \cite{WilsonIJCAI09} to allow a richer form of equivalence, generalising e.g., the $\modelsCoplustwo$ system,
then the preference inference will no longer be polynomial.

\commenta asdf
SOMETHING MORE ABOUT CONNECTION WITH \cite{WilsonECAI14} FORMALISM:
The formalism developed here and that described in \cite{WilsonECAI14} are more closely related than
might appear at first sight.
For each element of $c_i\in\calC$ we can generate a variable $X_i$,
and let $V$ be the set of these variables.
For each alternative $\alpha\in\calA$ we generate a complete assignment $\alpha^*$
on the variables $V$
by $\alpha*(X_i) = c_i(\alpha)$.
Note that values of $\alpha^*(X_i)$ are non-negative numbers,
and thus  have a fixed ordering, with zero being the best value.
Let us consider models in $\calCone$.
Each model $H \in \calCone$
corresponds to a sequence of evaluations, and thus
has an associated sequence of variables; this sequence together with the fixed value orderings,
generates a lexicographic order.
In contrast with the lexicographic inference system in \cite{WilsonECAI14},
the logic developed in this paper allows strict (as well as non-strict) preference statements,
and allows more than one variable at the same level.
However, the lexicographic inference logic from \cite{WilsonECAI14} does not assume a fixed value ordering

\commenta
**
More somewhere about the contributions of the paper.
Different systems at the end:
forms of lexicographic inference;
it is closely connected with a logic of disjunctive ordering statements.

\section*{Acknowledgments}
This publication has emanated from research conducted with the financial support of Science Foundation Ireland (SFI) under Grant Number SFI/12/RC/2289. Nic Wilson is also supported by the School of EEE\&CS, Queen's University Belfast.
We are grateful to the reviewers for their helpful comments.

\commenta
We are grateful for the reviewers' helpful comments.
OR
We are grateful to the reviewers for their helpful comments.


\bibliography{HCLP-logic}

\end{document}